\documentclass[12pt,twoside]{book}
\usepackage[dvips]{epsfig}
\usepackage{plenum}

\begin{document}

\chapter{HADRON INTERACTIONS--HADRON SIZES }

\author{Bogdan Povh}

\affiliation{Max--Planck--Institut f\"ur Kernphysik\\
Postfach 103980\\
D--69029 HEIDELBERG}


\section{INTRODUCTION}

Hadronic interactions are successfully parameterized in different
models. Those most frequently applied are the Regge parameterization and
the geometrical model. In the present paper we want to apply the
geometrical picture to different domains of the hadronic interaction. 
The model used is a very economical model in that it contains only
few parameters most of which have a direct physical interpretation.

We first want to apply the geometrical model to hadron--proton
interactions. Then we want to extend the model to deep inelastic
scattering. In deep inelastic scattering at small $x$, the
interaction can be viewed upon as an interaction of the hadronic
fluctuation of the photon with the proton. The size of the hadronic
fluctuation is $\langle{}r^2\rangle~\approx~1/Q^2$. Thus, deep
inelastic scattering offers a nice extension of the applicability of
the geometrical model for hadronic sizes from $1$\,fm down to
$10^{-3}$\,fm. However, the nuclear shadowing demonstrates that 
in addition to the ``hard'' interaction corresponding to a 
hadronic size of $1/Q^2$, an additional, ``soft'' component (large size)
of the interaction in deep inelastic scattering is
present. We will elaborate this in the section on nuclear structure
function. We will point out the importance of these results for the
theoretical treatment of the heavy ion reactions at high energies.


\section{HADRON--PROTON CROSS SECTION}

When analyzing the differential cross sections for high-energy $pp$
collisions available at that time, Wu and Yang\refnote{\cite{wu}}
as well as Chou and Yang\refnote{\cite{chou}}
observed that the $t$ dependence of the differential elastic cross
section is closely related to the charge form factor $F_i(t)~i=1,2$
of colliding hadrons. For small values of $t$ the form factors are
related to the mean-squared charge radii $\langle{}r^2_{ch}\rangle_i$
via

\begin{equation}
F_i(t)=1+\frac{1}{6}\langle r^2_{ch}\rangle_i t + 0(t^2)
\label{f1}
\end{equation}

\noindent 
which implies that the slope parameter $b_{12}$ is related
to the charge radii via

\begin{equation}
b_{12}=\frac{1}{3} \left(\langle{}r^2_{ch}\rangle_1 + 
                         \langle{}r^2_{ch}\rangle_2 \right). 
\label{f2}
\end{equation}

\begin{figure}[hbt]
\centering\epsfig{file=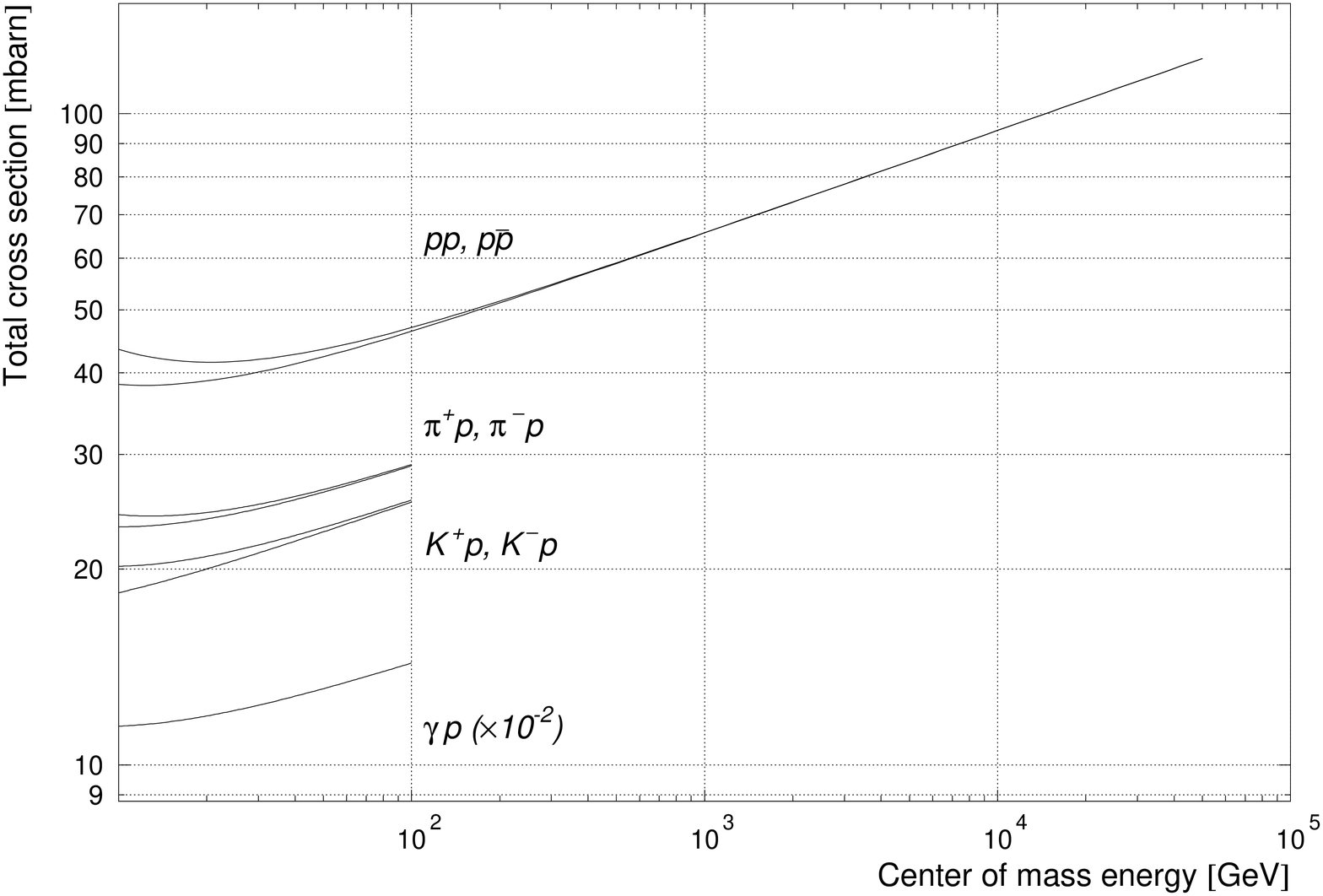, height=8.0cm}
\caption{Schematically shown the total cross sections of pions,
kaons, protons/antiprotons and photons on the proton.}
\label{data}
\end{figure}

\begin{figure}[hbt]
\centering
\begin{minipage}[b]{.4\linewidth}
\centering
\epsfig{file=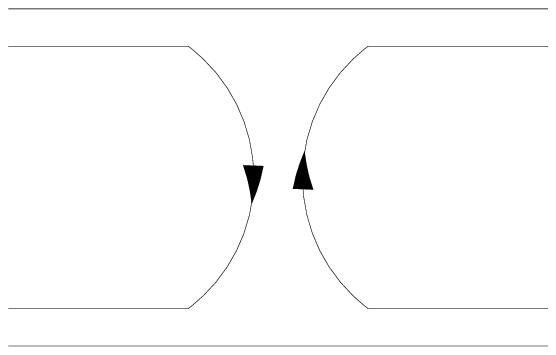}\\
{\footnotesize (a) Regge exchange}
\end{minipage}
\hspace*{.1\linewidth}
\begin{minipage}[b]{.4\linewidth}
\centering
\epsfig{file=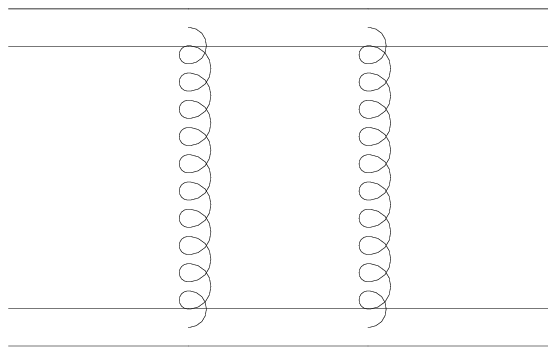}\\
{\footnotesize (b) Pomeron exchange}
\end{minipage}
\caption{Quark--antiquark or Regge exchange is schematically 
shown in figure~(a) while~(b) shows
the two gluon exchange or Pomeron exchange which is the 
dominant interaction at high energies.}
\label{regge}
\end{figure}

\begin{figure}[hbt]
\begin{minipage}[b]{0.38\linewidth}
\caption{The total cross section square is proportional to the 
imaginary part of the elastic scattering amplitude at zero degrees.}
\label{optical}
\end{minipage}
\hfill
\begin{minipage}[b]{0.6\linewidth}
\centering\vspace*{1.1cm}\epsfig{file=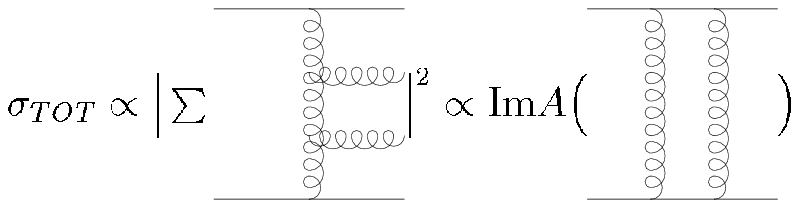}
\end{minipage}
\end{figure}

In a series of papers J. H\"ufner and I\refnote{\cite{povh}} have
shown additional regularities in the hadron--hadron cross sections. 
The experimental data are shown in Fig.~\ref{data}.
At small energies, $\sqrt{s}<10$\,GeV quark-antiquark (Regge) exchange
is the dominant interaction mechanism. The cross section falls off
with $\sigma_{TOT}\propto{}1/\sqrt s$ (see Fig.~\ref{regge}(a)). 
At energies $\sqrt{s}>10$\,GeV, the hadronic cross sections increase
logarithmically as well as the slope parameters do. In the Regge
nomenclature this regime is called Pomeron exchange
(see Fig.~\ref{regge}(b)).
At high energies the hadronic cross section is entirely absorptive!
The elastic cross section is therefore fully determined by absorption.
The total and the elastic cross section are connected via the optical
theorem symbolically as shown in Fig.~\ref{optical}.
Therefore in the geometrical model we consider the absorption and
treat elastic scattering as a shadow of it. The geometrical
properties of the hadronic interaction become transparent.

Introducing an effective hadron radius $R(s)$ the total cross section
can be written as

\begin{equation}
\sigma_{TOT}(s)=a R^2_1(s)R^2_2(s)
\label{f3}
\end{equation}

\noindent
and the slope parameter 

\begin{equation}
b_{12}(s)=\frac{1}{3}\left(R^2_1(s)+R^2_2(s)\right).
\label{f4}
\end{equation}

The constant $a$ is a universal constant, i.e. it is valid for all
hadrons and energies. Formula~(\ref{f3}) may look strange but we have
to remember that hadrons are color dipoles.

\noindent 
The energy dependence of the effective radius is given by

\begin{equation}
R^2(s)=R^2(s_0) \left( 1+\epsilon \ln \left(\frac{s}{s_0}\right) \right). 
\label{f5}
\end{equation} 

\begin{figure}[hbt]
\begin{minipage}[b]{.38\linewidth}
\caption{The hadron, a color dipole, is surrounded by a cloud of gluons.
The spreading of the gluons is governed by the random walk.}
\label{random}
\end{minipage}
\hfill
\begin{minipage}[b]{.6\linewidth}
\centering\epsfig{file=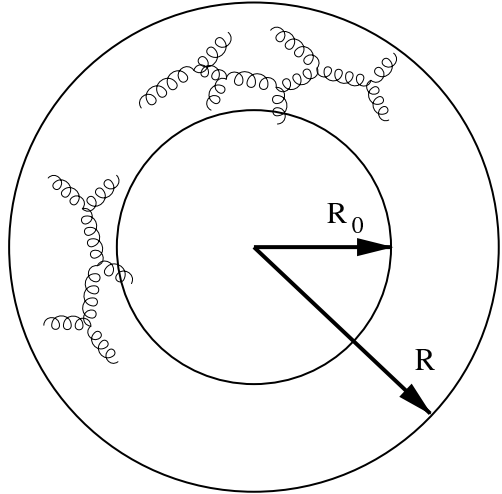}
\end{minipage}
\end{figure}

At small energies Regge exchange is the dominant mechanism and 
formula~(\ref{f5}) cannot be checked in this region. Nevertheless, 
we assume that
there is an effective hadron radius in the value close to the charge
radius. The logarithmic increase with energy (eq. [\ref{f5}]) can
been worked out perturbatively. We will,however, give only a
qualitative explanation of it. Figure~\ref{random} shows
the spreading of gluons beyond the border of the hadron
core.

Let us assume that at the radius $R_0$ a gluon has a Bjorken $x = x_0$. 
After $n$ splittings, for simplicity we assume in two equal parts,

\begin{equation}
x=\left(\frac{1}{2}\right)^n x_0.
\label{f6}
\end{equation}

\noindent 
The average distance of gluon traveling between splitting is $\lambda$.
Gluons spread according to the random walk, so

\begin{equation}
R^2-R^2_0=n \lambda^2 c \ln \frac{x_0}{x}.
\label{f7}
\end{equation} 

\noindent 
The interaction is totally absorptive, so the inelastic event is possible
only if

\begin{equation}
x_1x_2s \ge M^2_0
\label{f8}
\end{equation}

\noindent 
where $x_1$ and $x_2$ are Bjorken $x$ of hadron 1 and 2, respectively. 
Inserting the condition (\ref{f8}) into formula (\ref{f7}) gives (\ref{f5}).


\section{DEEP INELASTIC SCATTERING}

\begin{figure}[hbt]
\begin{minipage}[b]{.65\linewidth}
\centering\epsfig{file=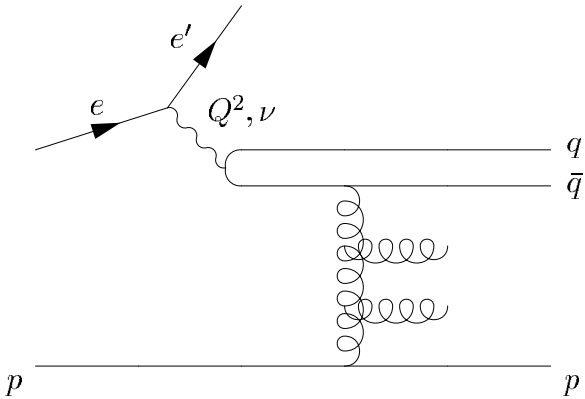}
\end{minipage}
\hfill
\begin{minipage}[b]{.33\linewidth}
\caption{Fluctuation of a virtual photon into a quark--antiquark pair.
The quark--antiquark pair interacts strongly with a proton.}
\label{quark}
\end{minipage}
\end{figure}

Deep inelastic scattering at small Bjorken $x$ can be viewed upon as
an interaction of a hadron fluctuation of the photon with the proton
(Fig. \ref{quark}). The fluctuation length of the $q\bar{q}$ pair is
given by the offshellness $\Delta E$ of the quark--antiquark system as
compared to the photon

\begin{equation}
\ell =\frac{1}{\Delta E} \approx \frac{1}{mx}
\label{f9}
\end{equation}

\noindent 
where $m$ is the proton mass and $x=Q^2/2m\nu$. One sees that
already for $x=0.1$ the fluctuation length is $\ell\approx1$\,fm and
the approximation of the interaction as a hadronic one is quite good.

In deep inelastic scattering one measures the following cross
section:

\begin{equation}
\frac{d\sigma}{dxdQ^2}= 
{\rm Photon flux}\,\cdot\,\sigma_{TOT}^{\gamma^\ast p}(x,Q^2)
\label{f10}
\end{equation}

\noindent 
where the photon flux is given by the probability of the electron to
emit a virtual photon which is believed to be well calculable within
the QED. The second factor is the total cross section for the
virtual photon $\gamma^\ast$ with the proton. The structure function
$F^p_2(x,Q^2)$, the main objective of deep inelastic scattering, is
related to $\sigma_{TOT}^{\gamma^\ast p}(x,Q^2)$ via

\begin{equation}
F_2(x,Q^2)\approx Q^2 \sigma_{TOT}^{\gamma^\ast p}(x,Q^2).
\label{structure}
\end{equation}

\noindent 
As pointed out above, for low $x$, the $\sigma^{\gamma^\ast p}_{TOT}$
can be expressed in terms of the hadron fluctuation (light cone
representation)

\begin{equation}
\sigma^{\gamma^\ast p}_{TOT}(x,Q^2)=\sum_h W_h(x,Q^2)\sigma_{TOT}(h,p).
\label{cone}
\end{equation}

\noindent 
The sum goes over all possible $q\bar{q}$ pairs. One could try to
calculate $W_h(x,Q^2)$ using some model of hadronic fluctuation. 
We will just assume this to be taken into account by the
normalization factor in the geometrical
cross section of the $q\bar{q}$ pair

\begin{equation}
\sigma_{TOT}(h,p) \alpha \left<r^2_{q\bar{q}}\right>= 
\frac{1}{m^2_q+\frac{Q^2}{4}}.
\label{radius}
\end{equation}

\noindent 
Here we assumed that for large $Q^2$ the transverse size of
$q\bar{q}$ pair is given by $1/Q^2$. For $Q^2\rightarrow 0$
confinement determines the size of the $q\bar{q}$ pair and the radius
of the hadronic fluctuation is just the radius of the $\rho$, the
well-known vector dominance regime. We see that eqs.~(\ref{radius})
and~(\ref{cone}) give automatically the Bjorken
scaling if inserted in eq.~(\ref{structure}).

\begin{figure}[hbt]
\centering\epsfig{file=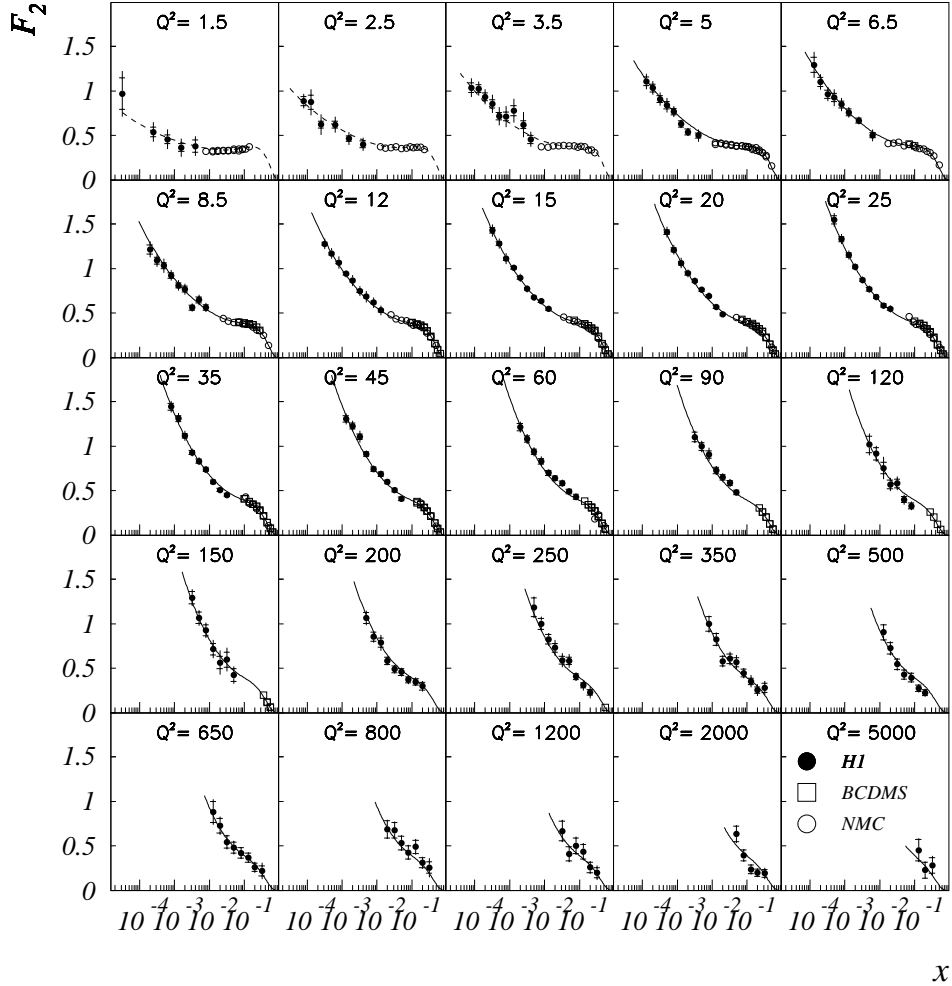, height=13.0cm}
\caption{Proton structure function $F_2(x,Q^2)$. Data are taken from
Ref.~\cite{h1}.}
\label{function}
\end{figure}

Conventionally, the total cross sections for virtual photon--photon
interaction are plotted as the structure function
(see~eq.~[\ref{structure}]). The measured structure functions as
published by the H1 collaboration\refnote{\cite{h1}} are shown in
Fig.~\ref{function}.

\noindent 
The Bjorken $x$ is given by $x=Q^2/2m \nu$ where $\nu$ is the energy
of the virtual photon. The nominator $2m \nu$ then is the square of
the center of mass energy of the photon--proton system

\begin{equation}
W^2=2m \nu.
\label{energy}
\end{equation}

\noindent 
Thus, for a fixed $Q^2$, $W^2\propto 1/x$. If we limit ourselves to
the $x$ domain in which the hadronic interpretation is meaningful,
the total $\gamma^\ast p$ cross sections are shown in Fig.~\ref{cross
section}.

\noindent 
The behavior of $\sigma_{TOT}(\gamma^\ast p)$ with the center of
mass energy square $W^2$ resembles strongly the one of the real
hadrons $\sigma_{TOT}(h,p)$. However, the increase of the cross 
section with energy is $Q^2$ dependent. For $Q^2\rightarrow 0$ it is
exactly that of the $pp$ and $p\bar{p}$.

We have explained the increase of the cross section with energy by
the increasing importance of the gluon halo in inelastic
processes. It seems, however, that we missed the $Q^2$ dependence in
this process. In fact, for the real hadrons, we considered in the
first section, the $Q^2\approx 0$ assumption is good enough. By
virtual hadrons with sizes $\approx1/Q^2$ the gluon halo starts with
gluons of higher and higher transverse momenta. We have to consider
the fact that the strong coupling constant $\alpha_s$ is $Q^2$
dependent.

This effect has been taken into account, for example in Ref.~\cite{DLLA}.
In the $Q^2$ region accessible to measurement now, the
$Q^2$ dependence of the gluon halo can be approximated by

\begin{eqnarray}
\sigma_{TOT}(Q^2,W^2) & = & a R^2_h(Q^2,W^2)R^2_p(Q^2=0,W^2) \nonumber\\
                      & = & \frac{A}{m^2_q+\frac{Q^2}{4}}
                            \left(1+\left(\varepsilon+\varepsilon'\ln
                            \left(1+\frac{Q^2}{Q^2_0}\right)\right)\ln
\frac{x_0}{x}\right) .
\label{fit} 
\end{eqnarray}

\begin{figure}[t]
\centering\epsfig{file=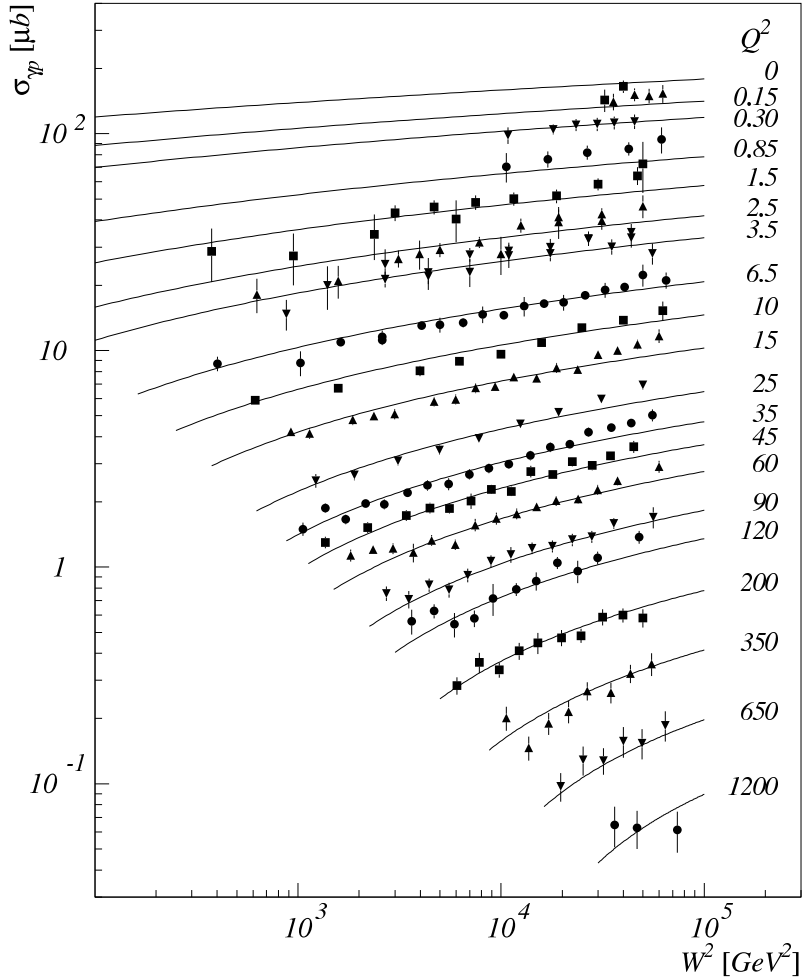, height=13.0cm}
\caption{Total $\gamma^\ast p$ cross sections deduced from the structure
functions of Fig.~\ref{function}. The cross sections are plotted only
for the energies where the geometrical model is applicable.}
\label{cross section}
\end{figure}
\noindent
Here $R^2_h$ is the radius square of the hadronic fluctuation, $R^2_p$
that of the proton and $a$ the coupling constant introduced in
eq.~(\ref{f3}). The curves in Fig.~(\ref{function}) are obtained by
applying eq.~(\ref{fit}).

Summarizing this section, one can conclude that the hadronic
interaction is well simulated by the interaction of two color
dipoles surrounded by a gluon cloud.


\section{NUCLEAR STRUCTURE FUNCTION}

The nuclear structure function is of interest because eventually
heavy ion reactions will be treated from ``first principles''. This
means that the quark and gluon structure functions have to be
used as input in the calculation of heavy ion reactions at
high energies. The collision is then described by elastic parton
scattering (see e.g. Ref. \cite{parton}). The nuclear structure function is
not just the sum of $A$ nucleon structure functions! As we show below,
nuclear shadowing distort the nuclear structure functions for
$x<0.1$. In order to obtain the gluon structure function of a nucleus
the evolution has to be applied to the measured quark structure
function. A direct determination of the quark nuclear structure
function is required. In this paper, however, we will consider the
nuclear structure function as a tool to improve our geometrical
picture of the virtual photon interaction.

\subsection{Nuclear shadowing}

\begin{figure}[hbt]
\centering\epsfig{file=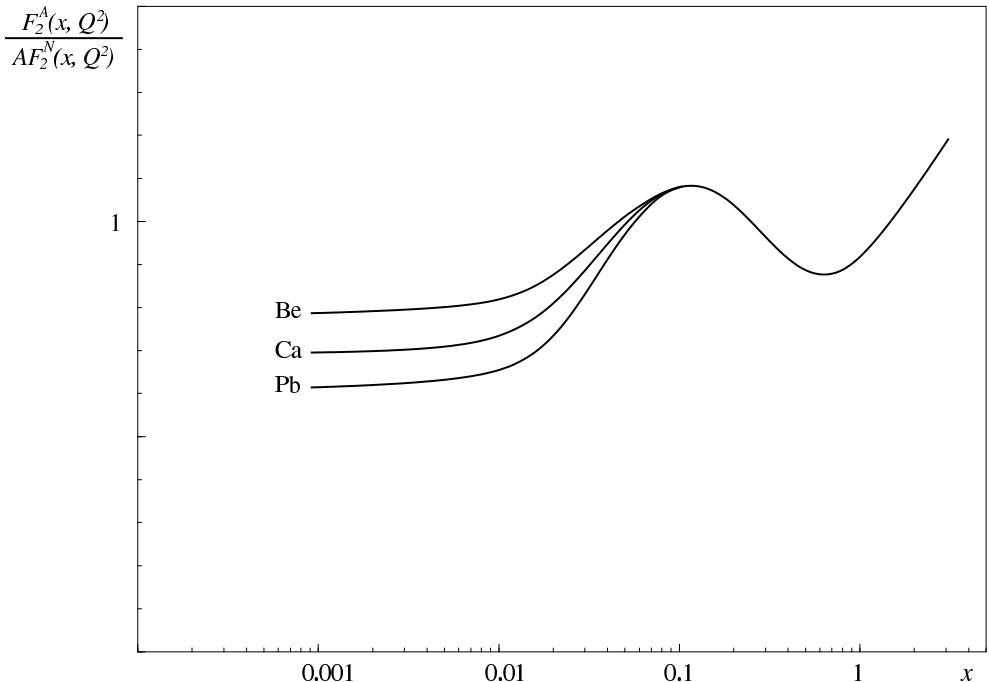, height=9.0cm}
\caption{Shadowing in nuclear structure functions shown schematically.}
\label{shadowing}
\end{figure}

Nuclear structure functions have extensively been measured by the NMC
collaboration\refnote{\cite{nmc}} at CERN in the eighties. 
Results of these measurements
are shown schematically in Fig.~\ref{shadowing}. We are interested in
the ratio of the nuclear structure function to the nucleon structure
function. At $x<0.1$ one observes shadowing, this means, the cross
section per nucleon is reduced, the reduction depending on the mass
number $A$. The surprising fact is that shadowing does not depend
strongly on $Q^2$.

Thus, the geometrical model we used above cannot be the full story.
If the cross section of the hadronic fluctuation goes down like
$1/Q^2$, then shadowing would very fast disappear according to

\begin{equation}
\frac{F^A_2(x,Q^2)}{AF^N_2(x,Q^2)}=1-\frac{\varepsilon}{Q^2}.
\label{disappear}
\end{equation}

\noindent 
As this is not the case, there must be a soft component of the
interaction. In fact, we have taken only a part of the geometrical
object presenting the $q\bar{q}$ pair. In our consideration, eq.~
(\ref{f3}), we assumed that the photon momentum is taken by quark and
antiquark equally. But this is not true.

\begin{figure}[t]
\begin{minipage}[b]{.48\linewidth}
\centering\epsfig{file=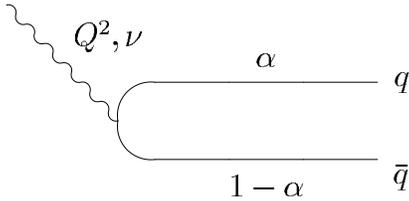}
\end{minipage}
\hfill
\begin{minipage}[b]{.5\linewidth}
\caption{Virtual photon fluctuates in $q\bar{q}$ pair. The quark takes
a fraction $\alpha$ of the photon momentum and the antiquark $1-\alpha$
or vice versa.}
\label{qqbar}
\end{minipage}
\end{figure}

\noindent
In Figure~\ref{qqbar} the quark takes over the ratio $\alpha$ of the
photon momentum and the antiquark $1-\alpha$ or vice versa.
Considering an asymmetric pair the radius square of the $q\bar{q}$
pair is (see Ref.~\cite{bjorkin})

\begin{equation}
\langle{}r^2\rangle \propto \frac{1}{m^2_q+\alpha(1-\alpha)Q^2}.
\label{alpha}
\end{equation}

\noindent 
For $\alpha=\frac{1}{2}$ we obtain the expression~(\ref{radius}) for
the radius square. For asymmetrical pairs $\alpha<1/Q^2$, the radius
of the $q\bar{q}$ is of $\rho$-meson size and the cross section
amounts to $\approx 20$\,mb. Certainly, the probability of having an
asymmetric pair with $\alpha<1/Q^2$ is small. The probability is
proportional to $1/Q^2$.

Now, dividing the interaction in a hard (small size probe),
corresponding to the approximately symmetric pairs, and a soft
component, corresponding to the pairs with $\alpha<1/Q^2$, we see
that both scale with $1/Q^2$ and cannot be distinguished in the
inclusive measurements of deep inelastic scattering on a proton.
On the nucleus, however, the contribution of the soft component is
strongly absorbed, dies off fast and is responsible for the
shadowing. For a quantitative treatment of shadowing see for example
Ref.~\cite{boris}.


\section{CONCLUSIONS}

A simple picture of a hadron and its interaction at high energies
emerges: The hadron is a color dipole of size $R$ surrounded by a
cloud of gluons. The interaction with the hadron core has to be
treated ``non-perturbatively'' with only parameters being the hadronic
size and the ``universal'' hadronic coupling constant.

The halo interaction can be understood within the perturbative
approach. This simple picture of the hadronic interaction is
supported by the results with hadrons of sizes $R\approx 1$\,fm down
to $R=0.01$\,fm corresponding to $Q^2\approx 2000$\,GeV$^2$ as
measured in deep inelastic scattering.



\end{document}